\documentclass[twocolumn,nofootinbib,amsmath,prl,aps,superscriptaddress,
tightenlines,preprintnumbers,nobalancelastpage]{revtex4-2}

\usepackage{slashed}
\usepackage{color, verbatim}

\usepackage{latexsym}
\usepackage{amsmath}
\usepackage{graphicx}

\usepackage{amssymb}
\usepackage{graphicx}
\usepackage{bm}
\usepackage{adjustbox}

\usepackage{graphicx}
\usepackage{latexsym}
\usepackage{amsmath}
\usepackage{mathrsfs}
\usepackage{enumerate}
\usepackage{adjustbox}
\usepackage{amssymb}
\usepackage{dsfont}      %
\usepackage{slashed}
\usepackage{graphicx}    %
\usepackage{wrapfig}     %
\usepackage{physics}
\usepackage{lipsum}

\def\bea{\begin{eqnarray}}
\def\eea{\end{eqnarray}}

\usepackage[normalem]{ulem}

\RequirePackage[hyperfootnotes=false,colorlinks=true,urlcolor=blue,anchorcolor=blue,citecolor=blue,filecolor=blue,linkcolor=blue,menucolor=blue,linktocpage=true,pdfproducer=medialab]{hyperref}

\hypersetup{
    colorlinks = true,
    linkcolor = blueCERN,
    anchorcolor = blueCERN,
    citecolor = blueCERN,
    filecolor = blueCERN,
    urlcolor = blueCERN
    }
\usepackage{orcidlink}
\definecolor{blueCERN}{HTML}{0033A0}
\usepackage{csquotes}
\definecolor{mygreen}{rgb}{0.0, 0.5, 0.0}

\usepackage[capitalise]{cleveref}
\crefname{equation}{Eq.}{Eqs.}
\Crefname{equation}{Eq.}{Eqs.}
\crefname{appendix}{App.}{Apps.}
\Crefname{appendix}{App.}{Apps.}

\usepackage{mathtools}
\usepackage{nicefrac}

\newcommand{\id}{\mathds{1}}   %

\definecolor{brightpink}{rgb}{1.0, 0.0, 0.5}

\usepackage{booktabs}
\begin{document}

\newcount\hour \newcount\minute
\hour=\time \divide \hour by 60
\minute=\time
\count99=\hour \multiply \count99 by -60 \advance \minute by \count99
\newcommand{\mydate}{\ \today \ - \number\hour :00}

\preprint{}

\title{\Large A Witten effect for global monopoles }

\author{Pablo Quílez Lasanta\orcidlink{0000-0002-4327-2706}}
\email{pablo.quilez.lasanta@cern.ch}
\affiliation{Theoretical Physics Department, CERN, 1 Esplanade des Particules, CH-1211 Geneva 23, Switzerland}

\preprint{CERN-TH/2026-248}

\begin{abstract}
Global monopoles can carry electric charge even when the scalar field that forms them is neutral.
We show that a CP-odd operator in the Goldstone effective theory, typically discarded as a total derivative,
gives them an electric charge $Q=-Cn$, with $C$ its Wilson coefficient and $n$ the
winding number.
This is a Witten effect on a global defect,
generated by a mixed $\theta$-term between
electromagnetism and a composite connection built from
the Goldstone fields.
The same group theory condition that allows the operator also guarantees the existence of monopoles.
We also give a fermionic UV completion and determine the coefficient
by matching the Goldstone--Wilczek current.

\end{abstract}

\maketitle

\newpage

\section{Introduction}

Total derivatives are often discarded from the Lagrangian, since they affect neither the classical equations of motion nor the perturbative S-matrix.
In theories with non-trivial topology, however, they can have physical consequences. The canonical example is the QCD $\theta$-term. Although it is a total derivative,  there exist non-trivial instanton configurations, classified by $\pi_3(SU(3))=\mathbb{Z}$, that render it physical~\cite{Belavin:1975fg,tHooft:1976rip,Callan:1976je,Jackiw:1976pf}.
Another example is the Witten effect~\cite{Witten:1979ey}. Electromagnetism has no instantons since $\pi_3(U(1))=0$, yet in the presence of other topological  defects, magnetic monopoles~\cite{tHooft:1974kcl,Polyakov:1974ek}, the QED $\theta$-term is observable since it induces an electric charge $Q=-e\,\theta_{\rm QED}/2\pi$ to the monopoles.
Induced quantum numbers on topological defects are a broader phenomenon, not restricted to gauge theories, \emph{e.g.} the Goldstone--Wilczek current~\cite{Goldstone:1981kk}.

Monopoles, in particular, do not need to arise from a gauge symmetry. The spontaneous symmetry breaking of a global $G\to H$ with $\pi_2(G/H)\neq 0$ gives rise to topological defects known as global monopoles~\cite{Barriola:1989hx,Vilenkin:2000jqa}. Unlike their gauged cousins, global monopoles carry no magnetic gauge charge and have no associated Dirac quantization condition~\cite{Dirac:1931kp,Zwanziger:1968rs}. Another difference is that their energy  grows linearly with the size of the system, but this does not make them unphysical. The Kibble mechanism~\cite{Kibble:1976sj} predicts their production in the early universe, where monopoles and antimonopoles form a network and annihilate efficiently~\cite{Bennett:1990xy}. For this reason they do not give rise to the \emph{monopole problem} of Grand Unified Theories~\cite{Preskill:1979zi}. Their annihilation radiates Goldstone bosons, which can contribute to the dark matter relic density~\cite{Nakano:2026scaling,Kanda:2026goldstone}. Belonging to a dark sector independent of QED, they are usually taken to be electrically neutral.

The same breaking pattern responsible for monopoles also generates Goldstone bosons (GB). These GB transforming under $H$ are known as darkly charged axion-like particles or \emph{DALPs} and have been studied in detail in Ref.~\cite{Arenaza:2026pxp} by deriving their general EFT with the CCWZ construction~\cite{Coleman:1969sm,Callan:1969sn}. In terms of the coset building block $u^i_\mu=\partial_\mu a_i/f+\mathcal{O}(a^2)$, where $a_i$ are the GB fields,
one of the operators with two Goldstone derivatives is
\begin{align}
    \mathcal{O}_{\tilde F a}= \frac{C}{4\pi}\,A_{ij}\, \widetilde{F}^{\mu\nu}\,u^i_\mu u^j_\nu ,
    \label{Eq:OperatorFmunudual}
\end{align}
where $\widetilde{F}^{\mu\nu}=\frac{1}{2}\epsilon^{\mu\nu\rho\sigma}F_{\rho\sigma}$ is the dual electromagnetic field
strength\footnote{Ref.~\cite{Arenaza:2026pxp} uses the hypercharge field strength $B_{\mu\nu}$.
We write it in terms of the photon, the hypercharge case follows with $eq\to g'Y$.} with $\epsilon^{0123}=1$, and $A_{ij}$ is an antisymmetric invariant tensor. We use a canonically normalized photon and absorb the electromagnetic coupling into $C$. Ref.~\cite{Arenaza:2026pxp} noted that the leading term in the Goldstone expansion of $\mathcal{O}_{\tilde F a}$ is a total derivative.

In this Letter, we show that this holds at all orders in $a/f$: the operator generates no local
interaction vertex and is therefore invisible in perturbation theory.
It is nevertheless physical in the presence of global monopoles. The reason is that
$\omega_{\mu\nu}\equiv A_{ij}u^i_\mu u^j_\nu/2\pi$ is the winding two-form of the Goldstone fields, so that
$\mathcal{O}_{\tilde F a}$ couples the winding of the defect directly to the electromagnetic field.
As a result, global monopoles acquire an electric charge
\begin{align}
    Q=- C\, n,
\end{align}
where $n$ is the winding number of the monopole. This is a Witten effect for global monopoles.

We identify the group theory condition behind this effect. Whenever the compact group $G$ is
semisimple and $H$ contains a $U(1)$ direct factor, $H=H'\times U(1)_H$, both ingredients
are present at once: \emph{i}) the operator $\mathcal{O}_{\tilde F a}$ is allowed, with the invariant
tensor $A_{ij}$ supplied by the generator of $U(1)_H$, and \emph{ii}) global monopoles exist, since
$\pi_2(G/H)\supseteq\mathbb{Z}$.
Therefore, within this class, the monopoles do not need to be postulated since the condition that permits the operator also guarantees the defects on which it acts.

The same $U(1)_H$ factor gives $\omega$ its geometric interpretation. This topological form
is the normalized field strength of the $U(1)_H$ connection built from the Goldstone fields, so that
$\mathcal{O}_{\tilde F a}$ can be understood as a mixed $\theta$-term in which one of the two
field strengths is composite rather than gauged.
Related topological portals involving Goldstone fields have been recently constructed
from mixed Wess--Zumino--Witten terms~\cite{Davighi:2024zip,Davighi:2024zjp}.

\section{Set-up}
\label{Sec:Setup}

Let us consider a global symmetry $G$ spontaneously broken to $H$. We assume $H$ to contain at least one $U(1)$ factor, $H=H'\times U(1)_H$, while the compact group $G$ does not, \emph{i.e.} $G$ is semisimple.

The breaking produces one Goldstone boson $a_i$ per broken generator $X_i$ and there are at least two
of them carrying opposite $U(1)_H$ charge. Following the CCWZ
construction~\cite{Coleman:1969sm,Callan:1969sn}, we define
\begin{align}
\xi = \exp\!\left[\, i\, \frac{a_i}{f_{i}}\, X_i \right],\qquad
\xi^{-1}\partial^\mu \xi = i\left( u^\mu + \Gamma^\mu \right),
\label{eq:ccwz}
\end{align}
where the Maurer--Cartan form $\xi^{-1}\partial^\mu\xi$ splits into the coset building block $u^\mu$,
along the broken generators, and the connection $\Gamma^\mu$, along the unbroken ones (see Ref.~\cite{Arenaza:2026pxp} for more details).

The $U(1)_H$ factor, whose generator is $T_H$, guarantees the existence of an antisymmetric $H$-invariant tensor
\begin{align}
A_{ij}=-i\,\Tr\!\left(T_H[X_i,X_j]\right),
    \label{Eq:Aij}
\end{align}
 and thus one can construct the operator $\mathcal{O}_{\tilde F a}$ of
Eq.~(\ref{Eq:OperatorFmunudual}), together with its $CP$-even partner built from $F^{\mu\nu}$ studied in Ref.~\cite{Arenaza:2026pxp}.

The same condition implies the existence of topological solitons. Static configurations of finite potential energy must approach the vacuum manifold at spatial infinity $S^2$, and thus they are classified by the second homotopy group of the coset space $\pi_2(G/H)$. Since $H$ contains a $U(1)$ factor while the compact semisimple $G$ does not, then
$\pi_2(G/H)\supseteq\mathbb{Z}$ (see \emph{e.g.}~\cite{Preskill:1984gd}), and the theory admits
global monopoles labelled by an integer winding number $n$.

\section{Global monopoles}
Let us consider the simplest breaking pattern within this class, $SU(2)\to U(1)$ where a triplet scalar field  $\varphi^a$ acquires a vacuum expectation value $\langle\varphi\rangle=v \, \sigma_3/2$. Since $\pi_2(SU(2)/U(1))=\mathbb{Z}$ the theory admits global monopoles. On the vacuum manifold, the winding two-form reads \cite{Arafune:1974uy,Tong:GaugeTheory}
\begin{align}
    &\omega_{\mu\nu}=\frac{1}{4\pi v^3}\epsilon_{abc}\varphi^a\partial_\mu \varphi^b \partial_\nu \varphi^c,&n=\int_{S^2} \omega \in \mathbb{Z}\,
    \label{Eq:windingCurrent}
\end{align}
Within this Goldstone description, $\omega$ is a closed 2-form $d\omega=0$ (see \cref{App:SU2}), so its integral over a closed 2-surface is unchanged by deformations of the surface that avoid the monopole core. This integral gives the winding (or monopole) number $n$ of the configuration,
which on a constant-time surface takes the familiar form~\cite{Tong:GaugeTheory}
\begin{align}
    n=\frac{1}{8\pi v^3}\int_{S^2} d^2S_i\;\epsilon^{ijk}\epsilon_{abc}\,
    \varphi^a\,\partial_j \varphi^b\,\partial_k \varphi^c\,.
    \label{Eq:winding_static}
\end{align}
The simplest $n=1$ configuration is the hedgehog solution
\begin{align}
    &\varphi^a=v\,h(r)\,\frac{{x}^a}{r}, & h(0)=0,\quad h(\infty)=1
    \label{Eq:hedgehog}
\end{align}
where $r=\sqrt{x^a x_a}$ is the 3D spatial radius. The radial profile $h(r)$ resolves the monopole core, where the field leaves the vacuum manifold and $d\omega=0$ no longer holds.
Substituting the hedgehog configuration in \cref{Eq:winding_static} one can check that it has winding number $n=1$. Here the monopole is a defect of the scalar fields and carries
no electromagnetic magnetic charge.

In the UV models of
Ref.~\cite{Arenaza:2026pxp} for this specific coset, it was already identified that the rhs of the first equation in \cref{Eq:windingCurrent} can be expressed as
$f_{ABC}\,\Phi_A\,\partial^\mu\Phi_B\,\partial^\nu\Phi_C\to A_{ij}\,u^{i\mu}u^{j\nu}$. This now implies that the winding two-form $\omega$ in \cref{Eq:windingCurrent} is exactly the form that couples to electromagnetism in the operator $\mathcal{O}_{\tilde F a}$ of \cref{Eq:OperatorFmunudual}.

This is not an accident of this example. Both $A_{ij}$ and $u^i_\mu$ are defined for an arbitrary
coset in the class of \cref{Sec:Setup}, and the winding two-form is
    \begin{align}
    &\omega_{\mu\nu}=\frac{1}{2\pi }A_{ij}u^i_\mu u^j_\nu\,,
    \label{Eq:windingCurrentUmu}
\end{align}
where $A_{ij}$ is the tensor of \cref{Eq:Aij}. The explicit factors use the $SU(2)/U(1)$ normalization, while for other cosets, $\omega$ needs to be normalized to integer periods
and absorb the group-dependent factor into $C$. The operator $\mathcal{O}_{\tilde F a}$ is then built from the object whose integral counts the winding of the Goldstone fields,
\begin{align}
    \mathcal{O}_{\tilde F a}= \frac{C}{2}\, \widetilde{F}^{\mu\nu}\, \omega_{\mu\nu}\,,
    \label{Eq:OperatorOmega}
\end{align}
or, in the language of differential forms,\footnote{The 2-form is
$\omega=\tfrac{1}{2}\omega_{\mu\nu}dx^\mu\wedge dx^\nu$. Since $A_{ij}u^i_\mu u^j_\nu$
is already antisymmetric in $\mu\nu$, the form $A_{ij}u^i\wedge u^j$ has components
$2A_{ij}u^i_\mu u^j_\nu$.}
\begin{align}
    S \supset \frac{A_{ij}}{4\pi} C\int F\wedge u^i \wedge u^j=C\int F\wedge \omega \,.
    \label{Eq:OperatorForm}
\end{align}
No metric appears in \cref{Eq:OperatorForm}, and the operator is therefore purely topological. By contrast, its
$CP$-even partner $\mathcal{O}_{Fa}\sim A_{ij}\,u^i_\mu u^j_\nu {F}^{\mu\nu}$ is not topological as it involves $F^{\mu\nu}\omega_{\mu\nu}$, \emph{i.e.} $F\wedge\star\,\omega$, which does
require the metric.

\section{A Witten effect}
\label{Sec:Witten}
We now show that the operator $\mathcal{O}_{\tilde F a}$ induces an electric charge on global
monopoles. We first solve for the electric field around a static hedgehog configuration in the simplest case,
$SU(2)\to U(1)$, and then generalize to arbitrary breaking patterns $G\to H$ and arbitrary winding
number $n$.

The winding two-form $\omega$ has antisymmetric components $\omega_{\mu\nu}$, which we can decompose
as we do for field strengths into an \emph{electric} $\mathcal{E}^i\equiv\omega^{0i}$ and a
\emph{magnetic} $W^k\equiv\tfrac{1}{2}\epsilon^{kij}\omega_{ij}$ component.
We stress that these labels are an analogy and $\vec{\mathcal{E}}$ and $\vec W$ are built
entirely from the Goldstone fields and are not electromagnetic fields.
 Time independence
ensures $\vec{\mathcal{E}}=0$,
so the
topological form of any static configuration is purely \emph{magnetic}. Specializing to the hedgehog solution of \cref{Eq:hedgehog} far away from the origin gives
\begin{align}
    \vec{W}=\frac{n}{4\pi}\frac{\vec{x}}{r^3}\,,\qquad \text{with }\, n=1\,.
    \label{Eq:Wfield}
\end{align}
Substituting it in the operator $\mathcal{O}_{\tilde F a}$, and using that for any two antisymmetric tensors
$\tfrac{1}{2}\widetilde{F}^{\mu\nu}\omega_{\mu\nu}=\vec{E}\cdot\vec{W}+\vec{B}\cdot\vec{\mathcal{E}}$,
\begin{align}
    \mathcal{O}_{\tilde F a}=C\left(\vec{E}\cdot\vec{W}+\vec{B}\cdot\vec{\mathcal{E}}\right)
    = C\,\vec{E}\cdot\vec{W}\,.
    \label{Eq:OperatorEW}
\end{align}

Now we can prove in two ways that the electric field around the monopole is the one generated by a
point charge. Following Coleman's derivation of the Witten effect~\cite{Coleman:1982cx}, we integrate the Lagrangian density over
space,
\begin{multline}
    L\supset C\!\int\! d^3x\;\vec{E}\cdot\vec{W}
    =-C\!\int\! d^3x\;(\vec{\nabla}A^0)\cdot\vec{W}\\
    =C\!\int\! d^3x\;A^0\,(\vec{\nabla}\cdot\vec{W})=C\,n\,A^0(0)\,,
    \label{Eq:ColemanRoute}
\end{multline}
where we used $\vec{E}=-\vec{\nabla}A^0$, integrated by parts\footnote{The surface term dropped here just
corresponds to the arbitrary constant value of the potential at infinity.} and applied
$\vec{\nabla}\cdot\vec{W}=n\delta^3(\vec{x})$ for the configuration in \cref{Eq:Wfield}.
Comparing with the coupling of a point charge, $\mathcal{L}\supset-\rho A^0$, we can obtain the electric charge of the monopole,
\begin{align}
    Q=-C\,n\,.
    \label{Eq:ChargeResult}
\end{align}

Alternatively, starting from the Lagrangian density,
\begin{align}
    \mathcal{L}&=-\frac{1}{4}F_{\mu\nu}F^{\mu\nu}+\mathcal{O}_{\tilde F a}
    =\frac{1}{2}\left(E^2-B^2\right)+C\,\vec{E}\cdot\vec{W}
\end{align}
one can compute the Euler--Lagrange equations. Varying with respect to
$A^0$ gives a modified Gauss law,
\begin{align}
    \vec{\nabla}\cdot\vec{E}=-C \, \vec{\nabla} \cdot \vec{W}=-C\,n\,\delta^3(\vec{x})\,.
\end{align}
Again, the monopole acts as a point charge $Q=-Cn$ localized at its core. This second derivation
involves no integration by parts and therefore no surface term at all.

For this derivation it was not crucial to specialize to $G=SU(2)$ or the
hedgehog, and it generalizes immediately to an arbitrary breaking pattern within the class of \cref{Sec:Setup}. For any static configuration the winding two-form is purely magnetic, so that
\cref{Eq:OperatorEW} holds. The Gauss law derived above $\vec\nabla\cdot\vec{E}=-C\,\vec\nabla\cdot\vec{W}$ is also general, and by integrating it over a volume $V$
enclosed by a closed surface $\Sigma$, we can obtain the electric charge $Q_\Sigma$,
\begin{multline}
    Q_\Sigma
    \equiv\int_V \!d^3x\,\vec\nabla\cdot\vec{E}
    =-C\oint_\Sigma \vec{W}\cdot d\vec{S}
    =-C\!\int_\Sigma\!\omega=-C\,n_\Sigma\,,
    \label{Eq:ChargeGeneral}
\end{multline}
where $n_\Sigma$ is the winding number of the configuration inside $\Sigma$. The electric
charge enclosed by a surface is therefore $-C$ times the winding through it.  It is rather remarkable that a field configuration of the Goldstone bosons, which are
electrically neutral, can carry an electric charge. But this topological phenomenon should not be unfamiliar, since the same is true of the Skyrmion, which carries baryon number despite being a configuration of pion fields that have a vanishing baryon number~\cite{Skyrme:1962vh,Witten:1983tx}. Two other closely related phenomena are the Jackiw--Rebbi effect~\cite{Jackiw:1975fn} where a kink in a scalar field can carry fermion number $1/2$, and the Goldstone--Wilczek current~\cite{Goldstone:1981kk} which gives kinks and magnetic monopoles fractional fermion number.

\section{Properties of the operator}
\label{Sec:Properties}

It may seem paradoxical that an operator which is a total derivative at every order in the Goldstone expansion, still gives the monopole a charge.
This phenomenon can be better understood through the geometry of $\omega$, which we now make explicit.

The connection built from the Maurer--Cartan form of \cref{eq:ccwz}, $\mathcal{A}_\mu\equiv -i\,\xi^{-1}\partial_\mu \xi = u_\mu + \Gamma_\mu$, is pure gauge
by construction, so its field strength $F(\mathcal{A})$ vanishes,
\begin{align}
    F(\mathcal{A})\equiv d\mathcal{A}+i\,\mathcal{A}\wedge\mathcal{A}=0\,.
    \label{Eq:MCflat}
\end{align}
This flatness condition is known as the Maurer--Cartan equation.
 One can project this field strength in the direction of $U(1)_H$ by tracing \cref{Eq:MCflat} against the corresponding generator $T_H$. This relates
the field strength of the $U(1)_H$ connection, $F^{U(1)_H}$, to the broken components
 $u_\mu$,
\begin{align}
    F_{\mu\nu}^{U(1)_H}=\partial_\mu \Gamma^{U(1)_H}_\nu-\partial_\nu \Gamma^{U(1)_H}_\mu
    =2A_{ij}u^i_\mu u^j_\nu = 4\pi\,\omega_{\mu\nu}\, ,
    \label{Eq:OmegaIsCurvature}
\end{align}
where $\Gamma^{U(1)_H}_\mu\equiv 2\,\Tr(T_H\Gamma_\mu)$ is the $U(1)_H$ component of the connection, and $\Tr (T_H^2)=1/2$.

The conclusion is that the winding two-form is the normalized field strength of the $U(1)_H$ connection built from the Goldstone fields $\omega=F^{U(1)_H}/(4\pi)$.
Unlike a true gauge field, however, $\Gamma^{U(1)_H}$ is not an independent degree of freedom. It
is instead fixed algebraically by the Goldstone configuration, which is why $U(1)_H$ has no propagating
photon.

As a consequence, $\mathcal{O}_{\tilde F a}$ can be understood as a mixed $\theta$-term,
\begin{align}
\mathcal{O}_{\tilde F a}
=\frac{C}{4\pi}\,A_{ij}\,\widetilde F^{\mu\nu}u^i_\mu u^j_\nu=\frac{e\bar\theta_{\rm mix}}{16\pi^2}
      \widetilde F^{\mu\nu}F^{U(1)_H}_{\mu\nu},
      \label{Eq:MixedTheta}
\end{align}
with $C=e\bar\theta_{\rm mix}/(2\pi)$.
Mixed $\theta$-terms between two gauge fields similarly induce electric
charges on hidden magnetic monopoles~\cite{Brummer:2009oul}.
Here, instead, one of the two field strengths is composite and made out of Goldstone fields rather than gauged. This explains why the
Witten effect also arises for global monopoles.

This identification also shows why the operator is invisible in perturbation theory. Using
the local relation in \cref{Eq:OmegaIsCurvature}, we obtain
\begin{align}
    \frac12\widetilde F^{\mu\nu}\omega_{\mu\nu}
    =\frac{1}{4\pi}\widetilde F^{\mu\nu}\partial_\mu\Gamma^{U(1)_H}_\nu
    =\frac{1}{4\pi}\partial_\mu\!\left(\widetilde F^{\mu\nu}\Gamma^{U(1)_H}_\nu\right),
    \label{Eq:TotalDerivative}
\end{align}
where we used the Bianchi identity $\partial_\mu\widetilde F^{\mu\nu}=0$.
Since at any finite order in $a/f$ the connection is a polynomial in the Goldstone fields,
the expansion of $\mathcal{O}_{\tilde F a}$ is a total derivative and it
generates no local vertex at any order.

However, the form $\omega$ is
not globally exact. The winding is non-trivial, $\int \omega=n$, precisely
because $\Gamma^{U(1)_H}$ is not globally defined.
This is the Dirac string at play, and it is most transparent in the explicit
$SU(2)\to U(1)$ example of \cref{App:SU2}.
Writing $a_1+i a_2=f\rho e^{i\psi}$, one finds
\begin{align}
    \Gamma^{U(1)}&=(1-\cos\rho)\,d\psi,\nonumber\\
    F^{U(1)}&=\sin\rho\,d\rho\wedge d\psi=4\pi\omega.
    \label{Eq:SU2Gamma}
\end{align}
where $d\rho=\partial_\mu\rho\,dx^\mu$ and $d\psi=\partial_\mu\psi\,dx^\mu$.
The connection is not well defined
at $\rho=\pi$, where all values of $\psi$ describe the same point.
For the hedgehog, this corresponds to the usual Dirac string along
the negative $z$-axis. The curvature  $F^{U(1)_H}$ is regular there, though, and it is only the
connection that requires at least two different charts. Thus
$F^{U(1)_H}=d\Gamma^{U(1)_H}$ only holds locally, without implying that the integral
$\oint_{S^2}F^{U(1)_H}$ vanishes. Instead, evaluating the flux gives $4\pi$ times the winding number (see \cref{Eq:App_stokes} in \cref{App:SU2} for the explicit integral)

Perturbation theory around $\rho=0$ stays within one regular chart,
so at every order in $a/f$ the operator is a total derivative and never contributes to perturbative amplitudes around a trivial vacuum.
A monopole configuration instead covers the full sphere and thus has access to the non-trivial topology of the vacuum manifold.
As shown at the end of \cref{App:SU2}, $\omega$ remains closed wherever
$\varphi\neq0$, even when its modulus varies. It is undefined at
the monopole centre, where the defect really lies with $\varphi=0$.

Another important property of this operator is that its Wilson coefficient depends on a periodic variable, an angle
$\bar\theta_{\rm mix}=2\pi C/e$, where $e$ is the smallest electric charge. How is periodicity of $\bar\theta_{\rm mix}$ compatible with the monopole inheriting a charge proportional to $C$?
A monopole of
winding $n$ can also carry any integer number $m$ of elementary charges (\emph{e.g.} by binding
charged particles), this gives the tower
 \begin{align}
Q_{m,n}
=e\left(m-\frac{n\bar\theta_{\rm mix}}{2\pi}\right),
\qquad m\in\mathbb Z,
\label{Eq:spectrum}
 \end{align}
 As for the original Witten effect, a shift $\bar\theta_{\rm mix}\to\bar\theta_{\rm mix}+2\pi$ changes the charge of each monopole by an integer,
 such that the whole spectrum remains invariant since the shift is absorbed by $m\to m+n$.

The operator is CP odd, so CP sends
$\bar\theta_{\rm mix}\to-\bar\theta_{\rm mix}$ and a state with winding $n$ and charge $q$ transforms $(n,q)\to(n,-q)$. Apart from the trivial  CP conserving point $\bar\theta_{\rm mix}=0$, for $\bar\theta_{\rm mix}=\pi$ the theory also conserves CP since the spectrum is invariant upon relabeling of $m$ in \cref{Eq:spectrum}.

Finally, in \cref{App:UV} we provide an explicit UV model where this operator
is generated by integrating out heavy
charged fermions. We consider an $SU(2)$ fermionic doublet
of electric charge $eq$, with a bare mass $M_F$ and a pseudoscalar
Yukawa coupling $y$ to the triplet $\varphi$. Matching the
Goldstone--Wilczek current~\cite{Goldstone:1981kk} to the electromagnetic
current of our operator we obtain,
\begin{align}
    \Delta\bar\theta_{\rm mix}=-q(\delta-\sin\delta),
    \qquad \delta=2\arctan\!\left(\frac{yv}{2M_F}\right),
\end{align}
where $\delta$ is the relative phase of the fermion mass eigenvalues.
\section{Conclusions and outlook}

We have shown that global monopoles can acquire electric charge through
a topological and CP-odd operator in the Goldstone EFT that is a total derivative at every order in the Goldstone
expansion. Its effect becomes physical in winding configurations,
giving the charge $Q=-Cn$. The operator couples electromagnetism with the winding two-form, which is the normalized field strength of a
composite $U(1)_H$ connection. This allows us to identify the operator as a mixed global-gauged
$\theta$-term.

This result follows directly from the Goldstone EFT, independently
of the microscopic origin of $C$. We identify the group theory conditions on the coset that ensure the existence of both global monopoles and the operator. Its Wilson coefficient is a periodic and CP-odd angle. We  also give a simple
fermionic UV completion, where matching the Goldstone--Wilczek current
determines the coefficient in terms of a relative phase in the mass matrix of the fermionic sector. This connects the EFT description of this work with the original Goldstone--Wilczek result of solitons acquiring
fractional fermion number.

The induced charges open several questions for the dynamics of global
monopoles. Monopole--antimonopole annihilation can produce pseudo-Goldstone dark matter~\cite{Nakano:2026scaling,Kanda:2026goldstone}. If this coupling is present, accelerating monopole--antimonopole pairs form time-dependent electric dipoles that emit electromagnetic radiation, which may affect their
evolution and annihilation and provide a possible observable signature.
It would also be interesting to study Goldstone-boson scattering off global monopoles
and the effects of both $\mathcal{O}_{\tilde F a}$ and $\mathcal{O}_{Fa}$ in monopole backgrounds.

Breaking patterns with several $U(1)$ factors in $H$ provide a natural extension,
with several winding charges and mixed $\theta$-terms. Other UV completions would
help clarify the relation between the induced coefficients and global-gauge anomalies, and the
role of additional contributions in the matching.

\vspace{1em}
{\bf Acknowledgments.}---%
I thank B.~Gavela, B.~Grinstein, E.~Jenkins, R.~Plestid, as well as the rest of the Porto
crew, and T. Steingasser for helpful discussions, and A.~Manohar for discussions and for valuable comments on an
early version of the manuscript. I thank the UCSD Physics Department and the Caltech group for
their hospitality, and I am also grateful for the inspiring views from the summit of Mont Buet
and the waves at Del Mar, where part of this work was thought through.
Claude Opus 5.5 and GPT-6 Astra assisted with language editing,
literature searches, and checks of conventions and normalization. I have verified all results presented in this manuscript. I acknowledge support from
the European Union's Horizon Europe research and innovation programme under the Marie
Sk\l odowska-Curie Postdoctoral Fellowship grant agreement No 101207780 (AxionCount) and the
Marie Sk\l odowska-Curie Staff Exchange grant agreement No 101086085 (ASYMMETRY).

\bibliographystyle{apsrev4-2}
\bibliography{refs}

\clearpage
\vspace{1em}

\onecolumngrid
\begin{center}
  \textbf{\large Supplemental Material}\\[0.3em]
  \textit{A Witten Effect for Global Monopoles}
\end{center}
\vspace{0.5em}
\twocolumngrid

\appendix
\setcounter{equation}{0}
\renewcommand{\theequation}{S\arabic{equation}}
\makeatletter\@removefromreset{equation}{section}\makeatother

\section{Explicit CCWZ expressions for $SU(2)\to U(1)$}
\label{App:SU2}

Let us consider the symmetry breaking pattern  $SU(2)\to U(1)_H$, where the broken generators are $X_i=\sigma_i/2$ ($i=1,2$), and $T_H=\sigma_3/2$ is the unbroken generator, all normalized so that
$\Tr(T_aT_b)=\delta_{ab}/2$.
Writing $\theta_i=a_i/f$ and $\theta_1+i\theta_2=\rho e^{i\psi}$, the coset
representative is
\begin{align}
    \xi&=e^{i\theta_iX_i}=\cos\frac{\rho}{2}\,\id
    +i\sin\frac{\rho}{2}(\sigma_1\cos\psi+\sigma_2\sin\psi).
\end{align}
The Maurer--Cartan form $\xi^{-1}d\xi=i(u^iX_i+\Gamma^3T_H)$ gives
\begin{align}
    u^i_\mu&=\frac{\sin\rho}{\rho}\partial_\mu\theta_i
    +\left(1-\frac{\sin\rho}{\rho}\right)
    \frac{\theta_i\theta_j}{\rho^2}\partial_\mu\theta_j,\nonumber\\
    \Gamma^3_\mu&=\frac{1-\cos\rho}{\rho^2}
    (\theta_1\partial_\mu\theta_2-\theta_2\partial_\mu\theta_1).
    \label{Eq:App_exact}
\end{align}
where $\Gamma^3=\Gamma^{U(1)_H}$ is the
full connection, since $H$ only contains the abelian factor $H=U(1)_H$.
These expressions are exact, \emph{i.e.} they hold to all orders in the GBs. In terms of the polar
variables $(\rho,\psi)$, and in form notation, $u^i=u^i_\mu dx^\mu$ and $\Gamma^3=\Gamma^3_\mu dx^\mu$,
they read
\begin{align}
    u^1&=\cos\psi\,d\rho-\sin\psi\sin\rho\,d\psi,\nonumber\\
    u^2&=\sin\psi\,d\rho+\cos\psi\sin\rho\,d\psi,\nonumber\\
    \Gamma^3&=(1-\cos\rho)\,d\psi.
    \label{Eq:App_polar}
\end{align}

The scalar $\varphi=\varphi^a\sigma_a/2=|\varphi|\,\xi T_H\xi^{-1}$ is then parametrized as
\begin{align}
    \frac{\varphi^a}{|\varphi|}
    =(-\sin\rho\sin\psi,\,\sin\rho\cos\psi,\,\cos\rho),
    \label{Eq:App_scalar_direction}
\end{align}
with $0\leq\rho\leq\pi$ and
$0\leq\psi<2\pi$.
For a hedgehog configuration, $\varphi^a=v\,h(r)\,x^a/r$ with $h(0)=0$ and $h(\infty)=1$, the
direction of $\varphi$ in field space coincides with the direction of $x^a$ in 3D space,
\begin{align}
    \frac{\varphi^a}{|\varphi|}\bigg|_{\rm hedgehog}=\frac{x^a}{r}
    =(\sin\vartheta\cos\phi,\,\sin\vartheta\sin\phi,\,\cos\vartheta)\,,
    \label{Eq:App_hedgehog}
\end{align}
for $r>0$, where $\vartheta$ and $\phi$ are the usual polar and azimuthal angles. Comparing
with \cref{Eq:App_scalar_direction} the hedgehog configuration corresponds to the following map between the spatial angles and the coset angles:
\begin{align}
    \rho&=\vartheta,\qquad \psi=\phi-\frac{\pi}{2},
    \label{Eq:App_hedgehog_map_angle}
\end{align}
The map to the Goldstone coordinates reads
\begin{align}
    a_1&=f\vartheta\sin\phi,\qquad a_2=-f\vartheta\cos\phi.
    \label{Eq:App_hedgehog_mapGB}
\end{align}
Now we are ready to build our winding two-form $\omega$. The antisymmetric invariant reads
$A_{12}=-A_{21}=-i\Tr(T_H[X_1,X_2])=-\frac{i}{8}\Tr\!\left(\sigma_3[\sigma_1,\sigma_2]\right)=\frac12$,and thus
\begin{align}
    4\pi\omega=u^1\wedge u^2=\sin\rho\,d\rho\wedge d\psi=d\Gamma^3=F^{U(1)}.
    \label{Eq:App_curvature}
\end{align}
 In components,
$4\pi\omega_{\mu\nu}=2A_{ij}u^i_\mu u^j_\nu$, as in the main text.
Interestingly, $\omega$ is nothing but the area form $\omega=\sin\vartheta\,d\vartheta\wedge d\phi/(4\pi)$ and thus the winding number coincides with the area of the normalized unit sphere: $\int_{S^2}\omega=1$.

Even though $4\pi\omega=d\Gamma^3$ locally, its integral over the closed sphere
does not vanish because $\Gamma^3$ is not globally defined. At $\rho=\pi$,
all values of $\psi$ describe the same point, while $\Gamma^3$
is not well defined there. As usual, parametrization of the sphere with the Euler angles
requires two charts: if we expand around the north pole, then the south pole is not well defined. Instead of using 2 charts, we can still compute the winding number by removing a small cap around the south pole and applying Stokes' theorem to the remaining region. The boundary
integral of $\Gamma^3=(1-\cos\rho)d\psi$ then gives
\begin{align}
    n=\int_{S^2}\omega
    =\lim_{\rho\to\pi^-}\frac{1-\cos\rho}{2}=1.
    \label{Eq:App_stokes}
\end{align}

This apparent singularity of $\Gamma^3$ along the negative $z$ axis is the usual Dirac string. It depends, however, on the choice of chart, while the winding form $\omega$ is actually regular there. This becomes explicit when we write $\omega$ directly in terms of the scalar field, allowing its modulus to vary:
\begin{align}
    \omega=\frac{1}
    {8\pi|\varphi|^3}\,\epsilon_{abc}\,\varphi^a\,d\varphi^b\wedge d\varphi^c.
    \label{Eq:App_omega_scalar}
\end{align}
It agrees with \cref{Eq:windingCurrent} at $|\varphi|=v$ and is defined
wherever $\varphi\neq0$. In order to check that $d\omega=0$  we will use the identity
\begin{align*}
    &\epsilon_{abc}\,d\varphi^a\wedge d\varphi^b\wedge d\varphi^c
    =\frac{3}{|\varphi|}\,d|\varphi|\wedge
    \epsilon_{abc}\varphi^a\,d\varphi^b\wedge d\varphi^c.
\end{align*}
Differentiating \cref{Eq:App_omega_scalar} then yields
\begin{align*}
    d\omega&=\frac{\epsilon_{abc}\,d\varphi^a\wedge d\varphi^b\wedge d\varphi^c}
    {8\pi|\varphi|^3}
    -\frac{3\,d|\varphi|\wedge\epsilon_{abc}\varphi^a\,d\varphi^b\wedge d\varphi^c}
    {8\pi|\varphi|^4}=0.
\end{align*}
Thus $\omega$ is closed wherever $\varphi\neq0$. The zeros of $\varphi$, where $\omega$ is
ill-defined, are precisely the monopole cores, which act as a point sources allowing a nonzero flux $\oint_{S^2}\omega=n$ through any sphere enclosing them.

\section{An example UV completion}
\label{App:UV}
In this appendix, we give a simple UV completion and determine the contribution
to the coefficient of $\mathcal{O}_{\tilde F a}$ from the phases of a heavy-fermion mass matrix.

Let us consider the symmetry breaking pattern of \cref{App:SU2}, $SU(2)\to U(1)_H$, realised by an
adjoint (triplet) scalar $\varphi$ with $\langle\varphi\rangle=v\,T_H=v\,\sigma_3/2$. The matter content is a
single Dirac fermion $\Psi$ in the fundamental representation, carrying electromagnetic charge $eq$,
where $q$ is dimensionless.

The fermion has a real bare mass $M_F>0$ and a real pseudoscalar Yukawa coupling $y$ to
$\varphi$,
\begin{align}
    \mathcal{L}\supset
    \bar\Psi\, i\gamma^\mu\!\left(\partial_\mu+ieqA_\mu\right)\Psi
    -\bar\Psi\big(M_F\,\id+i y\,\gamma_5\varphi\big)\Psi .
    \label{Eq:UVlagGW}
\end{align}
 Upon symmetry breaking, the mass matrix reads
\begin{align}
    M_\Psi = M_F\,\id + i y |\varphi|\,T_H
    =\begin{pmatrix} M_F+\frac{i y|\varphi|}{2} & 0\\[3pt] 0 & M_F-\frac{i y|\varphi|}{2}\end{pmatrix} .
    \label{Eq:MmGW}
\end{align}
Note that this is not the most general mass term, but instead a particular choice that allows us
to use directly the one-loop result for the Goldstone--Wilczek (GW) current in Ref.~\cite{Goldstone:1981kk}.
Indeed, in this model, the electromagnetic current is given by the GW fermion number
current \cite{Goldstone:1981kk} multiplied by $eq$ and thus reads,
\begin{align}
    j^\mu_{\rm em,UV}
    &=\frac{eqM_Fy^3}{6\pi^2\left(4 M_F^2+y^2|\varphi|^2\right)^2}\epsilon^{\mu\nu\rho\sigma}\epsilon_{abc}
    \partial_\nu\varphi^a\partial_\rho\varphi^b\partial_\sigma\varphi^c\,.
    \label{Eq:GWCurrentUV}
\end{align}
It is useful to introduce the relative phase $\delta$ between the two fermion mass eigenvalues,
\begin{align}
    \delta(|\varphi|)
    &\equiv\arg\!\left[\frac{M_F+iy|\varphi|/2}{M_F-iy|\varphi|/2}\right]=2\arctan\!\left(\frac{y|\varphi|}{2M_F}\right).
    \label{Eq:GWRelativePhase}
\end{align}

In terms of $\delta$, the current reads
\begin{align}
    j^\mu_{\rm em,UV}
    =\frac{eq}{48\pi^2}\,\frac{1-\cos\delta}{|\varphi|^2}\,\frac{d\delta}{d|\varphi|}\,
    \epsilon^{\mu\nu\rho\sigma}\epsilon_{abc}
    \partial_\nu\varphi^a\partial_\rho\varphi^b\partial_\sigma\varphi^c\,.
    \label{Eq:GWCurrentPhase}
\end{align}

We determine the heavy fermion contribution to our operator by matching this current to the
electromagnetic current in the effective theory.
For this purpose, we do allow the coefficient $C=C(|\varphi|)$ to depend on a slowly varying background $|\varphi|\neq 0$. Only after matching we set $|\varphi|=v$.

Using integration by parts we can rewrite $\mathcal{O}_{\tilde F a}$ as\footnote{We drop the total derivative, since it does not contribute to the local current. Note that the whole operator $\propto C(|\varphi|) F\wedge\omega$ is no longer a total derivative for varying $|\varphi|$.}
\begin{align}
    \frac{C(|\varphi|)}2\widetilde F^{\mu\nu}\omega_{\mu\nu}
    =\frac12 A_\mu\epsilon^{\mu\nu\rho\sigma}
    \partial_\nu\!\left[C(|\varphi|)\omega_{\rho\sigma}\right]
    \,.
    \label{Eq:GWOperatorIBP}
\end{align}
 With the convention $-A_\mu j^\mu_{\rm em}$,
we therefore obtain
\begin{align}
    j^\mu_{\rm em,IR}
    =-\frac{\delta S_{\rm mix}}{\delta A_\mu}
    =-\frac12\epsilon^{\mu\nu\rho\sigma}
    \partial_\nu\!\left[C(|\varphi|)\omega_{\rho\sigma}\right].
    \label{Eq:GWCurrentIRVariation}
\end{align}
where $S_{\rm mix}=\int d^4x\,\mathcal{O}_{\tilde F a}$. As shown in \cref{App:SU2}, $d\omega=0$ wherever $\varphi\neq0$, including for configurations
with a varying modulus. Consequently, the derivative in \cref{Eq:GWCurrentIRVariation} acts only on
$C(|\varphi|)$, and using the chain rule together with some Levi-Civita gymnastics we obtain
\begin{align}
    j^\mu_{\rm em,IR}
    &=-\frac{1}{24\pi|\varphi|^2}\frac{dC}{d|\varphi|}\epsilon^{\mu\nu\rho\sigma}\epsilon_{abc}
    \partial_\nu\varphi^a\partial_\rho\varphi^b\partial_\sigma\varphi^c
    \label{Eq:GWCurrentIR}
\end{align}
This has exactly the same tensor structure as the GW UV current in \cref{Eq:GWCurrentUV}. Matching the coefficients we obtain
\begin{align}
    \frac{dC}{d|\varphi|}
    &=-\frac{eqM_Fy^3|\varphi|^2}
    {4\pi\left(M_F^2+y^2|\varphi|^2/4\right)^2},
    \label{Eq:GWMatchingModulus}
\end{align}
or, equivalently,
\begin{align}
    \frac{dC}{d\delta}=-\frac{eq}{2\pi}(1-\cos\delta)\,.
    \label{Eq:GWMatchingPhase}
\end{align}
Integrating and setting $|\varphi|=v$ we obtain the Wilson coefficient in the Goldstone EFT,
\begin{align}
    C=-\frac{eq}{2\pi}\left[\delta-\sin\delta\right].
    \label{Eq:GWVacuumCoefficient}
\end{align}
where the integration constant is fixed by $C(|\varphi|=0)=0$.
Recalling that $\mathcal{O}_{\tilde F a}$ is a mixed $\theta$-term, \cref{Eq:MixedTheta}, with
$C=e\bar\theta_{\rm mix}/(2\pi)$, the contribution of the heavy fermions to the mixed $\theta$-angle reads
\begin{align}
    \Delta\theta_{\rm mix}=-q\left(\delta-\sin\delta\right).
    \label{Eq:GWInducedOperator}
\end{align}
The same function of $\delta$ gives the fractional baryon number of chiral bags~\cite{Goldstone:1983tu}.

For $M_F\gg|y|v$, one has $\delta\simeq yv/M_F$ and \cref{Eq:GWInducedOperator} reduces to
\begin{align}
    \Delta\theta_{\rm mix}\simeq-\frac{q}{6}\left(\frac{yv}{M_F}\right)^3,
    \label{Eq:GWHeavyLimit}
\end{align}
which is compatible with the CP-odd coefficient obtained in
Ref.~\cite{Arenaza:2026pxp} from a direct one-loop computation
expanded in the Yukawa coupling, upon identifying their Yukawa coupling with $-iy$ and
$g_1Y_F$ with $eq$.

For $M_F\to0$ in \cref{Eq:GWInducedOperator} one finds $\delta\to\pi$ for $y>0$, so that for $q=1$ we obtain
\begin{align}
    \bar\theta_{\rm mix}\to-\pi\equiv\pi\pmod{2\pi}\,,\qquad C\to-\frac e2\,,
    \label{Eq:GWHalfCharge}
\end{align}
and thus a monopole with unit winding number then carries charge $e/2$. This is the
electrically charged global monopole counterpart of the fermion number $\tfrac12$ found in the same limit by
Goldstone and Wilczek~\cite{Goldstone:1981kk}, and of the Jackiw--Rebbi
result~\cite{Jackiw:1975fn}. In this limit the theory is CP invariant, since
$-\pi\equiv\pi$, and a nonzero monopole charge is consistent with this: CP maps the tower
onto itself.

When the fermion masses lie below $v$, the same result can alternatively be obtained with the following procedure. One first builds the intermediate EFT between $v$ and $M_\Psi$, in which the
dressed fermion $\Psi'=\xi^{-1}\Psi$ couples, through
its covariant derivative, to both the electromagnetic potential and the composite connection
$\Gamma^{U(1)_H}_\mu$ (as well as to the coset building blocks $u^i_\mu$). Making the mass matrix
real and positive then requires a chiral rotation that has a
$U(1)_{\rm em}\times U(1)_H$ anomaly. Since this rotation does not commute with the $u^i_\mu$
couplings, the corresponding anomaly not only yields the term linear in $\delta$ in \cref{Eq:GWInducedOperator}, but also the $\sin\delta$ term due to the interactions with the Goldstones~\cite{Wess:1971yu,Witten:1983tw}. Indeed, for the case of gauge magnetic monopoles, the anomaly would only give $\Delta\bar\theta_{\rm mix}=-q \delta$, as outlined in \cite{Goldstone:1981kk}. For general mass matrices, which can not be written as \cref{Eq:MmGW}, one must carefully take into account all terms in the fermion effective action, as computed for the general case in Refs.~\cite{Manohar:1984anomalous,Salcedo:2000chiral}.

\newpage

 \newpage

\end{document}